\documentclass[12pt,preprint,url]{aastex}
\usepackage{amsbsy}
\usepackage{amsmath}
\usepackage{rotating}
\newcommand{\be}{\begin{equation}}
\newcommand{\ee}{\end{equation}}

\newcommand{\igr}{IGR J00291$+$5934}
\newcommand{\xte}{XTE J1751$-$305}
\newcommand{\sax}{SAX J1808.4$-$3658}
\newcommand{\swi}{Swift J1756.9$-$2508}
\newcommand{\rstar}{R_\ast}
\newcommand{\mstar}{M_\ast}
\newcommand{\msun}{M_\odot}

\def\ltsima{$\; \buildrel < \over \sim \;$}

\def\lsim{\lower.5ex\hbox{\ltsima}}
\def\gtsima{$\; \buildrel > \over \sim \;$}
\def\gsim{\lower.5ex\hbox{\gtsima}}
\shorttitle{Spin down XXX}
\shortauthors{Melatos}
\begin{document}
\title{Electromagnetic spin down of a transient accreting millisecond pulsar during quiescence}

\author{A. Melatos\altaffilmark{1}, A. Mastrano\altaffilmark{1}}

\email{amelatos@unimelb.edu.au, alpham@unimelb.edu.au}

\altaffiltext{1}{School of Physics, University of Melbourne,
Parkville, VIC 3010, Australia}

\begin{abstract}
\noindent
The measured spin-down rates in quiescence of the transient accreting millisecond pulsars {\igr}, {\xte}, {\sax}, and {\swi} have been used to estimate the magnetic moments of these objects assuming standard magnetic dipole braking. It is shown that this approach leads to an overestimate, {if} the amount of residual accretion is enough to distort the magnetosphere away from a force-free configuration{, through magnetospheric mass loading or crushing}, so that the lever arm of the braking torque migrates inside the light cylinder. We derive an alternative spin-down formula {and} calculate the residual accretion rates where the formula is applicable. {As a demonstration, we apply the alternative spin-down formula to produce updated magnetic moment estimates for the four objects above. We note that, based on current uncertain observations of quiescent accretion rates, magnetospheric mass loading and crushing are neither firmly indicated nor ruled out in these four objects. Because quiescent accretion rates are not measured directly (only upper limits are placed), it is impossible to be confident without more data whether the thresholds for magnetospheric mass loading or crushing are reached or not.}
\end{abstract}


\section{Introduction
 \label{sec:qui1}}
Four accreting millisecond pulsars (MSPs), the transient systems {\igr}, {\xte}, {\sax}, and {\swi}, are observed to spin down during the quiescent interval between accretion episodes \citep{hetal08,p10,pam10,retal11}. Hitherto,
the deceleration has been interpreted as arising from a standard magnetic dipole torque, just as in isolated, rotation-powered pulsars \citep{mel97,spi06,con06}. Under this interpretation, the measured spin-down rate can be inverted to infer the dipole magnetic field strength at the stellar surface, $B_\ast=2\mu/\rstar^3$, where $\mu$ and $\rstar$ denote the magnetic dipole moment and stellar radius respectively. The available X-ray timing data imply $0.9 \lesssim B_\ast/10^8\,{\rm G} \lesssim 6$ for the above four objects, consistent with the recycling scenario in general and magneto-centrifugal spin up in particular \citep{wij98}.

Magnetic dipole braking, as traditionally conceived, requires a rotation-powered pulsar to have a properly developed, {force-free,} electron-positron magnetosphere extending out to the light cylinder at cylindrical radius $R_{\rm L} = c/\Omega$, where $\Omega$ is the angular speed. {Under force-free conditions \citep{m91,b10},} the light cylinder {coincides with} the lever arm, where the stellar magnetic field lines are swept back, and the electromagnetic torque is effectively exerted. \footnote{The lever arm has length $R_{\rm L}$, whether the dipole is point-like \citep{ost69} or extended \citep{mel97}, but the spin-down law is modified in the latter case.} In this context, `properly developed' means that the magnetosphere hosts
exactly the right charge and current distributions to sustain the so-called `oblique rotator solution', {supplied} by electron-positron pairs created in vacuum gap cascades near the polar cap and/or $R_\textrm{L}$ \citep{m96,spi06}. It is doubtful {that this structure can be sustained over the long term} in an accreting system. Even during quiescence, residual high-density plasma from the accretion process is expected to {leak into the magnetosphere, until force-free conditions cease to apply} \citep{ill75,che85,luo07,cor08}. A neutron star with a mass-loaded magnetosphere still spins down, of course, but {the inertial forces are significant \citep{che85}, and the lever arm is effectively the Alfv\'{e}n radius $R_\textrm{A}<R_\textrm{L}$ rather than $R_\textrm{L}$ \citep{betal06}.} In other words, the size of the corotating magnetosphere (and hence the lever arm) is set by the ram pressure of the residual accretion flow from the previous accretion episode, not {by the self-consistent conduction and displacement currents in the force-free solution. Mass loading} is also expected to switch off the pulsar radio emission, {by shorting out (`poisoning') the parallel electric fields which power the vacuum gap cascades, but poisoning is a separate physical process which does not affect the braking torque directly \citep{che85,cor08}.}

In this paper, we quantify the maximum residual accretion rate that can be tolerated, before {the force-free approximation breaks down and inertial effects become important} in an accreting environment during quiescence. We consider two mechanisms, which modify the magnetic dipole braking torque away from its standard form: {magnetospheric mass loading (\S\ref{sec:qui2})} and `crushing' by accretion ram pressure (\S\ref{sec:qui3}). The results are compared with theoretical and observational limits on the residual accretion rate from a remnant disk during quiescence in \S\ref{sec:qui5}. {It is found that current observationally-inferred upper limits of residual accretion rates do not rule out conclusively magnetospheric mass loading or crushing in four particular accreting pulsars (IGR J00291$+$5934, XTE J1751$-$305, SAX J1808.4$-$3658, and Swift J1756.9$-$2508)}. In \S\ref{sec:qui6}, we present an alternative braking formula, which should be used when the standard magnetic dipole picture cannot be applied. We reanalyze the data from {\igr}, {\xte}, {\sax}, and {\swi} to provide updated limits on $B_\ast$.

\section{Magnetospheric mass loading
 \label{sec:qui2}}

{A rotation-powered pulsar magnetosphere is force-free, provided that two conditions are met \citep{m91,b10}: (i) the mechanical energy density is much less than the electromagnetic energy density; and (ii) the charge density equals} the Goldreich-Julian value required to sustain the motional electric field, $2\epsilon_0 {\bf\Omega} \cdot {\bf B}$, where $\bf{\Omega}$ is the angular velocity, and ${\bf B}$ is the local magnetic field strength, {everywhere except in thin `vacuum gaps'} in the inner and/or outer magnetosphere. {Without accretion, pair production in the vacuum gaps guarantees that there is enough plasma to satisfy condition (ii) in the systems of interest here. The force-free magnetosphere carries a self-consistent,} relativistic conduction current density $2\epsilon_0 {\bf{\Omega}}\cdot{\bf{B}}c$, which is comparable to the displacement current density at $R_\textrm{L}$ and spins down the star via a magnetic dipole braking torque $\propto B_\ast^2 \rstar^6 R_\textrm{N}^{-4}\Omega^{-1}\propto B_\ast^2 \rstar^6 \Omega^3${, with effective lever arm $R_\textrm{N}=R_\textrm{L}$ \citep{betal06}.}

{The dominance of electromagnetic stresses [condition (i)] can be expressed in terms of the magnetization parameter \citep{m91}}

\begin{equation} \sigma = \frac{e B_\ast\rstar^3\Omega^2}{4 m_e c^3} \left(\frac{n}{n_\textrm{GJ}}\right)^{-1},\end{equation}
{where $n$ denotes the plasma number density, and $n_\textrm{GJ}=2\epsilon_0 {\bf\Omega} \cdot {\bf B}/e$ is the Goldreich-Julian value. Force-free conditions apply for}

\be \frac{\sigma}{\gamma} > 1,\ee
{where $\gamma$ is the Lorentz factor of the gap-accelerated magnetospheric plasma. The left-hand side of equation (2) equals $cB^2/\mu_0 n \gamma m c^3$, the ratio of the Poynting flux to the mechanical energy flux, up to a factor of order unity. Given the standard scaling $\sigma/\gamma \propto r^{-3}$ as a function of radius $r<R_\textrm{L}$, with $n\propto B \propto r^{-3}$ and $\gamma\approx$ constant, it is enough to have $\sigma/\gamma > 1$ at $r=R_\textrm{L}$ in order for the magnetosphere to be force-free everywhere in the region $r<R_\textrm{L}$.}

{Now suppose that} residual accretion loads the magnetosphere all the way down to the surface with excess plasma, {with number density $n_a$. Three-dimensional simulations show that leakage into the magnetosphere is facilitated by tongue-like accretion streams and two-stream instabilities, which occur even when the system is in the magnetocentrifugal (propeller) regime \citep{rkl08}. The rate of leakage increases with magnetic obliquity \citep{retal03}. Equations (1) and (2) then imply that the force-free approximation breaks down, and inertial forces become important, for}

\be \frac{n_a}{n_\textrm{GJ}(R_\textrm{L})} > \frac{e B_* \Omega^2\rstar^3}{4\gamma m_e c^3}.\ee
{Making the standard simplifying assumption \citep{gl79}} that residual accretion occurs spherically at roughly the free-fall speed, $v \approx (GM_\ast/r)^{1/2}$ (where $\mstar$ is the neutron star mass), we can relate the accretion rate $\dot{M}_{\rm a}$ (mass per unit time) to the number density at radius $r$ approximately by

\begin{equation}
 \dot{M}_{\rm a}
 = 4\pi r^2 m_p n_{\rm a}(r) (GM_\ast/r)^{1/2},
\label{eq:qui1}
\end{equation}
where $m_p$ is the proton mass.\footnote{{A more realistic (and difficult) calculation involving tongue-like streams implies higher $n_a$ and more mass loading locally, so equation (4) is conservative from the perspective of force-free breakdown.}} Then the maximum residual accretion rate that can be tolerated before {force-free conditions break down, $\dot{M}_\textrm{ff}$,} is given by

\begin{eqnarray}
 \frac{\dot{M}_{\rm ff}}{\dot{M}_{\rm E}}
 & = &
 \frac{\epsilon_0  B_\ast^2 \rstar \sigma_{\rm T}}{2\gamma m_e}\left(\frac{R_\textrm{L}}{\rstar}\right)^{-9/2}
 \left(
  \frac{GM_\ast}{\rstar c^2}
 \right)^{-1/2}
\label{eq:qui3}
 \\
 & = &
 1.6\times 10^{-10}
 \left(
  \frac{\Omega}{10^3 \, {\rm rad\,s^{-1}}}
 \right)^{9/2}
 \left(
  \frac{B_\ast}{10^8\,{\rm G}}
 \right)^2\left(\frac{\gamma}{10^6}\right)^{-1},
\label{eq:qui4}
\end{eqnarray}
where $\sigma_{\rm T}$ is the Thomson cross-section, equation (\ref{eq:qui4}) follows from (\ref{eq:qui3}) for the canonical values $M_\ast = 1.4 M_\odot$ and $\rstar = 10\,{\rm km}$, and we normalize by the Eddington rate with unit radiative efficiency, $\dot{M}_{\rm E} = 4\pi GM_\ast m_p/(c\sigma_{\rm T})$. Equation (\ref{eq:qui4}) defines a relatively low accretion rate. We ask how low, in terms of plausible models of residual accretion, in \S\ref{sec:qui5}.

{The Lorentz factor $\gamma$ in equation (6) depends on the extent to which the parallel electric fields inside the vacuum gaps are shorted out. In turn, this depends on the detailed physics of the instabilities and diffusion processes controlling mass loading and is difficult to predict from first principles. If poisoning is effective, one obtains $\gamma\approx 1$, i.e., weak acceleration like in the equatorial `dead zone' \citep{luo07}. If poisoning is ineffective, one obtains $\gamma\approx \gamma_b/\kappa$, where $\gamma_b\approx 10^7$ is the Lorentz factor of the primary beam, and $\kappa$ is the multiplicity of the pair production process, with $10\lesssim \kappa \lesssim 10^4$ for inverse-Compton- and curvature-triggered cascades \citep{hib01}. We do not express a preference for either regime here and leave the $\gamma^{-1}$ scaling in (6) for the reader's convenience. We emphasize that gap poisoning enters the problem only in this indirect sense, through its effect on $\gamma$ and hence $\dot{M}_\textrm{ff}$; it does not modify the spin-down torque directly. We mention for completeness that poisoning can conceivably work in the opposite sense too. \citet{mit90} and \citet{mit91} predicted that an interstellar comet passing through a dead pulsar's magnetosphere may short-circuit the outer gaps and initiate a transient pair cascade, which may trigger a gamma-ray burst.}

\section{Magnetospheric crushing
 \label{sec:qui3}}
The ram pressure of the residual accretion can also disrupt the operation of a rotation-powered pulsar by crushing its magnetosphere. Let the Alfv\'{e}n radius $R_{\rm A}$ be the distance, where the electromagnetic momentum flux in the magnetosphere stands off the accretion flow. There are three crushing scenarios: (i) $R_{\rm A} > R_{\rm L}$: the classical, force-free magnetosphere is undisturbed, the electromagnetic torque is exerted at $R_{\rm L}$, and the standard magnetic dipole braking formula applies; (ii) $\rstar < R_{\rm A} < R_{\rm L}$: there is still an undisturbed portion of the magnetosphere, just above the stellar surface, where vacuum gaps can form, and a Poynting-flux-dominated wind is launched, but the outer magnetosphere (including at $r\approx R_\textrm{L}$) is distorted away from its normal structure, so the effective lever arm and hence the magnetic dipole braking formula are modified; and (iii) $R_{\rm A} = \rstar$: the magnetosphere is completely disrupted, and the object cannot function as a rotation-powered pulsar.



Typically $R_{\rm A}$ is determined by balancing the magnetic pressure of a static dipole against the ram pressure of matter falling at the free-fall speed, viz.

\be R_\textrm{A}=\mu^{4/7}(G\mstar)^{-1/7}\dot{M}_\textrm{a}^{-2/7},\label{eqra}\ee
up to a factor of order unity \citep{gl79,hgc11,mp14}. To crush the magnetosphere, at least partially, one requires $R_\textrm{A}< R_\textrm{L}$ [scenarios (ii) or (iii) above], which occurs when the residual accretion rate $\dot{M}_\textrm{a}$ exceeds a threshold $\dot{M}_\textrm{L}$ given from equation (\ref{eqra}) by

\begin{eqnarray}
 \frac{\dot{M}_{\rm L}}{\dot{M}_{\rm E}}
 & = &
 \frac{\epsilon_0 B_\ast^2 \rstar \sigma_{\rm T}}{4 m_p}
 \left(
  \frac{R_{\rm L}}{\rstar}
 \right)^{-7/2}
 \left(
  \frac{GM_\ast}{\rstar c^2}
 \right)^{-3/2}
\label{eq:qui6}
 \\
 & = &
 6.3\times 10^{-6}
 \left(
  \frac{\Omega}{10^3 \, {\rm rad\,s^{-1}}}
 \right)^{7/2}
 \left(
  \frac{B_\ast}{10^8\,{\rm G}}
 \right)^2~.
\label{eq:qui7}
\end{eqnarray}

The question of whether the magnetosphere is crushed can be asked in another, related way: is the pulsar's spin-down luminosity $L_{\rm SD}$ (carried predominantly by a Poynting-flux-dominated wind beyond $R_\textrm{L}$) high enough to blow away the residual accretion flow? The momentum flux transported by the wind, $L_{\rm SD}/(4\pi r^2 c)$, scales $\propto r^{-2}$ for $r>R_\textrm{L}$, whereas the accretion ram pressure scales $\propto r^{-5/2}$. Hence, the wind blows away the accreting gas everywhere beyond $R_{\rm L}$, provided that $L_{\rm SD}/(4\pi r^2 c)$ exceeds the accretion ram pressure at $R_{\rm L}$. The threshold accretion rate in this scenario, $\dot{M}_\textrm{SD}$, satisfies $\dot{M}_\textrm{SD}/\dot{M}_\textrm{L}\sim 1$ and is a function of magnetic inclination angle as in standard oblique rotator formulae \citep{spi06,con06}.

{In general, the threshold for crushing (\ref{eq:qui7}) is higher than that for mass loading (\ref{eq:qui4}). The reader may question how there can be enough matter in the magnetosphere for mass loading if the crushing threshold is not exceeded. In other words, a system where the accretion rate is lower than the threshold given by (\ref{eq:qui7}) should be in the `propeller' state, and all matter should be ejected from the system. However, this is an open question. The simulations of \citet{retal03}, \citet{retal04}, and \citet{rkl08} showed that matter is still accreted towards the star along `tongues' that penetrate the magnetosphere. Kelvin-Helmholtz and Rayleigh-Taylor instabilities, as well as collisional cross-field diffusion (neglected in MHD models), allow leakage into the magnetosphere. In fact, any magnetic field line which connects the disk and the star can be a conduit, just like for aurorae on Earth and Jupiter. The aforementioned simulations are in a slightly different parameter regime, so it is hard to be certain what happens, but mass loading without crushing is plausible.}

\section{Residual accretion
 \label{sec:qui5}}
The next task is to estimate the rate $\dot{M}_\textrm{a}$ at which residual accretion occurs in an accreting MSP during periods of quiescence. This comes down to examining theoretical
predictions of, and observational upper limits on, the presence of a remnant disk. If $\dot{M}_\textrm{a}$ exceeds either $\dot{M}_\textrm{ff}$ or $\dot{M}_\textrm{L}$, then the standard magnetic dipole braking law is modified.

Theoretically, residual accretion disks should persist around accreting MSPs during quiescence for some of the same physical reasons that debris disks persist around isolated neutron stars after supernova fallback . The constituent asteroids of a neutron star's fallback disk can undergo collisional migration, whereby inelastic collisions between the asteroids tend to broaden the disk and transport mass inwards towards the neutron star, and Yarkovsky migration, whereby uneven heating of the asteroids by the neutron star tends to shrink the disk \citep{r98,cor08}. The characteristic lifetime of a fallback disk undergoing collisional migration is $\sim 3$ Myr (proportional to the disk's mass, density, radius, size of the asteroids, the root-mean-square speed of the asteroids, and the neutron star's spin frequency), whereas the characteristic lifetime of a fallback disk undergoing Yarkovsky migration is $\sim 10^{3.1}$ Myr [proportional to the disk radius, the size of the asteroids, and inversely proportional to the neutron star's luminosity, the inner radius of the disk, and the constituent asteroids' drift rate ($\approx 10^{-3}$ AU Myr$^{-1}$ near the light cylinder of a typical pulsar)] \citep{cor08}.

In addition, numerical simulations show that the accretion disk around a neutron star can be in one of three regimes: (1) $\dot{M}_\textrm{a}/\dot{M}_\textrm{co}\gg 1$, (2) $\dot{M}_\textrm{a}/\dot{M}_\textrm{co}\approx 1$, or (3) $\dot{M}_\textrm{a}/\dot{M}_\textrm{co}\approx 0$, where $\dot{M}_\textrm{co}=\eta\mu^2/(\Omega R_\textrm{co}^5)$ is the accretion rate that puts the inner edge of the disk at $R_\textrm{co}=(G\mstar/\Omega^2)^{1/3}$, and $\eta\lesssim 0.1$ is the ratio between the azimuthal magnetic field strength (generated by star-disk differential rotation) and the poloidal magnetic field strength at the inner edge of the disk \citep{das10,das12}. In regime (1), the neutron star accretes and spins up until the inner edge of the disk approaches $R_\textrm{co}$ and the system enters regime (2). In regime (2), mass is prevented from accreting by a centrifugal barrier, but may not gain enough speed to be flung out of the system \citep{ss77,st93}. If $\dot{M}_\textrm{a}$ continues to fall, the system enters regime (3); if, on the other hand, mass piles up in the inner regions of the disk until it overcomes the centrifugal barrier, accretion restarts, and the disk radius approaches $R_\textrm{co}$ \citep{das10,das11,das12}. Within regime (2), the disk can vacillate between accreting and non-accreting states, or it can become unstable, depending on $\dot{M}_\textrm{a}$ and the depth to which the star's magnetic field penetrates the disk \citep{das10}. The disk often gets `trapped' just outside $R_\textrm{co}$, even as $R_\textrm{co}$ moves outwards as the star spins down, because $\dot{M}_\textrm{a}$ is too low \citep{das12}.

Fallback debris disks have been confirmed around neutron stars, for example around the magnetar 4U 0142$+$61 \citep{wck06} [but see also \citet{wetal08} for an alternative explanation] and, more spectacularly, as a planetary system around the radio pulsar PSR B1257$+$12 \citep{wf92}. Recent searches for disks around other radio pulsars have not been successful \citep{wetal14}. Once detected, optical and infrared (IR) spectra can be used to infer the temperature and hence the inner radius of the disk, although the latter inference relies on knowing the distance and albedo \citep{wck06}. The debris disk of 4U 0142$+$61, for example, has an inner temperature (inferred from the shape of the near-IR spectrum) that is comparable to the sublimation temperature of dust, suggesting that the inner radius of this disk may be set by X-ray destruction of dust, although the possibility that the radius might have been set by past interactions with the magnetosphere cannot be ruled out \citep{wck06}. Gas can continue to flow inwards through the sublimation radius and accrete. Circumbinary disks have also been detected in IR/near-IR around some low-mass X-ray binaries \citep{mm06,ww14}. These disks may be remnants of fallback debris disks or may consist of matter lost from the low-mass companion \citep{mm06}.

{However, note that unambiguous evidence of the existence of accretion disks around AMSPs [e.g., double-peaked emission lines of the Balmer series \citep{zetal14}] has not been reported. According to radio ejection models \citep{bur03,cetal04}, a disk should not exist at all.} Direct measurements of residual accretion rates for the four MSPs discussed in this paper are {therefore impossible to set}.


In {\sax}, \citet{hom01} proposed that the optical flux during quiescence can be used to infer a residual accretion rate $\dot{M}_\textrm{a,obs}\approx 9\times 10^{-12} (M_2/0.05\phantom{i}\msun)^2$ $\msun$ yr$^{-1}$, where $M_2$ is the companion's mass. {However, Burderi et al. (2003), Campana et al. (2004), and Deloye et al. (2008) showed that, {in order to explain the amplitude of the modulation of the optical flux at the orbital period}, the donor star must be irradiated by an external energy flux two orders of magnitude greater than the measured X-ray luminosity. These authors suggested that the irradiation comes from the spin-down power output of SAX 1808.4$-$3658 itself. {Furthermore, \citet{del08}, \citet{wetal09}, and \citet{wetal13} modeled the light curves, including the contribution of a putative residual accretion disk, and found that the disk contributes $\lesssim 30\%$ of the total optical emission. In another AMSP, XTE J1814$-$338, \citet{detal09} and \citet{betal13} also found that a residual accretion disk contributes $\lesssim 20\%$ of the optical flux. Because of the difficulty of deriving tight constraints on $\dot{M}_\textrm{a,obs}$ from optical flux measurements, we do not use this method.}

An upper limit on the mass accretion rate of SAX J1808.4$-$3658 has been evaluated by Heinke et al. (2009) at $9\times 10^{-12} \msun$ yr$^{-1}$, {by averaging over outbursts and quiescence over 12 yr,} which presupposes that there is no mass loading or crushing. {Alternatively, \citet{wetal13} fitted the disk contribution to the optical emission of {\sax} with a disk extending down to the light cylinder radius to obtain a temperature profile $T(r)\approx 6.2\times 10^{3} (r/10^{8} \textrm{ m})^{-1/2} \textrm{ K}$. At $r=R_\textrm{L}$, this gives $T=1.8\times 10^5$ K. Using the relation valid for a Shakura-Sunyaev accretion disk $T_\textrm{in}^4=3GM\dot{M}_\textrm{a}/(8\pi\sigma R_\textrm{in}^3)$ \citep{ss73}, where $R_\textrm{in}$ and $T_\textrm{in}$  are the truncation radius of the accretion disk and the temperature at the truncation radius, we find an accretion rate of $\approx 7\times 10^{-14} \msun$ yr$^{-1}$. In the quiescent state, the Shakura-Sunyaev solution does not hold and the accretion rate may be higher. To be conservative, we use the upper limits given by Heinke et al. (2009) (based on an average rate over outbursts and quiescence) in Table 1.}


{Heinke et al. (2009) analysed X-ray data from {\xte} and obtained $\dot{M}_\textrm{a,obs}\approx 6\times 10^{-12}$ $\msun$ yr$^{-1}$ as an upper limit (an average over outbursts and quiescence)}, although no optical spectrum has been detected to provide an independent check \citep{jetal03,detal09}. Deep quiescent monitoring of {\swi} has not been attempted \citep{hetal09b}, but \citet{pam10} used fluence data from its 2007 and 2009 outbursts to estimate an average accretion rate of $\dot{M}_\textrm{a,obs}=1.5\times 10^{-11}$ $\msun$ yr$^{-1}$.

Optical/IR spectra can be inconclusive. The optical/IR emission from {\igr} can be explained by irradiation of its companion; it cannot be used to infer anything about the disk and $\dot{M}_\textrm{a}$ \citep{detal07}. An upper limit of $\dot{M}_\textrm{a,obs}\lesssim 2.5\times 10^{-12}$ $\msun$ yr$^{-1}$ has been set by fitting the X-ray spectrum with a thermal component (radiated by the neutron star) and a power-law component (due to the quiescent accretion), but the evidence for the thermal component is only at the $3\sigma$ level, and the distance is poorly known \citep{hetal09b}. Assuming a distance of 4 kpc, the knee-like feature in the light curve during {\igr}'s outbursts implies an outburst accretion rate of $2\times 10^{-10}$ $\msun$ yr$^{-1}$ up to a factor of $\sim 2$ (distance uncertainty) \citep{hgc11}. {The value $\dot{M}_\textrm{a,obs}=2.5\times 10^{-12}$ $\msun$ yr$^{-1}$ is obtained as an average over outbursts and quiescence and hence should be understood as an upper limit on the quiescent accretion rate.} {The X-ray flux is $\sim 10^3$ times larger during an outburst than during quiescence \citep{hetal09b}, therefore the long-term average can be biased strongly towards the outburst value. In particular, if the mass transfer is not conservative and/or the disk is not completely emptied during an outburst, the average X-ray flux cannot be expected to trace the average mass accretion rate anymore. In this paper, for definiteness, we assume that the mass transfer is conservative and the average mass accretion rate can be inferred from the average X-ray flux, subject to the above strong caveats.}

{A recent paper by Mukherjee et al. (2015) estimated the accretion rate and surface field strengths for 14 MSPs, including the four MSPs discussed in this paper, independently from the arguments presented the previous paragraphs. They inferred the accretion rate from the lowest and highest observed X-ray fluxes during epochs that exhibit pulsations. We take the lowest fluxes exhibiting pulsations as the upper limit of the quiescent accretion rate and list them in the final column of Table 1. In Section 6 below, we compare our approach to that of Mukherjee et al. (2015) briefly.}

We normalise the above values of $\dot{M}_\textrm{a,obs}$ by the Eddington accretion rate, $\dot{M}_\textrm{E}=3.1\times 10^{-9}$ $\msun$ yr$^{-1}$ (for $\mstar=1.4$ $\msun$), and list them in Table \ref{table2}. We use the values as a consistency check against those predicted by the modified braking model in \S\ref{sec:qui6}.

{As discussed in \S 5, we assume that $R_\textrm{in}$ are close to $R_\textrm{co}$. The previously-mentioned Shakura-Sunyaev relation implies $\sigma T_\textrm{in}^4\propto \dot{M}_\textrm{a}/R_\textrm{in}^3$. Therefore, for a given $\dot{M}_\textrm{a}$, a smaller $R_\textrm{in}$ implies a greater X-ray emission flux. An estimate for the mass inflow luminosity as $G\mstar\dot{M}_\textrm{a}/R_\textrm{in}$ gives, in the case of {\sax}, for $\dot{M}_\textrm{a}\sim 10^{-11} \msun$ yr$^{-1}$, a luminosity of $\sim 10^{32}$ erg s$^{-1}$, which is consistent with observations. According to a more careful analysis by \citet{cetal98}, one needs $\dot{M}_\textrm{a}\sim 10^{-13}\msun$ yr$^{-1}$ to yield a mass inflow luminosity of $\sim 10^{34}$ erg s$^{-1}$ (for a neutron star with spin period 4 ms and field strength of $10^8$ G). However, this updated estimate of $\dot{M}_\textrm{a}$ is still higher than the thresholds for magnetospheric mass loading, as discussed in \S 5 below. Furthermore, note that these values of $\dot{M}_\textrm{a}$ assumes perfect radiative efficiency, which is optimistic. We stress that $\dot{M}_\textrm{a,obs}$ presented in Table 1 are (rather lax) upper limits, not direct measurements. }

\begin{table*}
\centering
 \begin{minipage}{140mm}

  \caption{Observational {upper limits} from X-ray spectra of residual accretion rates $\dot{M}_\textrm{a,obs}$ for four accreting MSPs, normalised to $\dot{M}_\textrm{E}$. {The second and third columns are derived by the methods summarized in Section 4 and the minimum pulsed X-ray flux respectively (Mukherjee et al. 2015).}}
  \begin{tabular}{lcc}
  \hline
    Name & $\dot{M}_\textrm{a,obs}/\dot{M}_{\textrm{E}}$ $(10^{-3})$ & $\dot{M}_\textrm{a,obs}/\dot{M}_{\textrm{E}}$ $(10^{-3})$\\
     & (X-ray flux) & (pulsations)\\
\hline
IGR J00291+5934 & $<0.8$ & $<3.2$\\
XTE J1751$-$305 & $<1.9$ & $<63$\\
SAX J1808.4$-$3658 & $<2.9$ & $<1.1$\\
Swift J1756.9$-$2508 & $<4.8$ & $<41$\\
\hline
\end{tabular}
\end{minipage}
\label{table2}
\end{table*}

\section{Modified braking torque
 \label{sec:qui6}}

We see from Table 1 that $\dot{M}_\textrm{a,obs}$ is higher than $\dot{M}_\textrm{ff}$ (for $\gamma\gtrsim 10^{2}$) for all four transient accreting MSPs and is higher than $\dot{M}_\textrm{L}$ for {\swi} (see Table 2). When the Mukherjee et al. (2015) estimate is used, $\dot{M}_\textrm{a,obs}$ is higher than $\dot{M}_\textrm{ff}$ (for $\gamma\gtrsim 10$) for all four transient accreting MSPs and is higher than $\dot{M}_\textrm{L}$ for {\swi} and {\xte}, with {\igr} and {\sax} on the borderline. {The upper limits of quiescent accretion rates quoted in Sec. 4 suggest that magnetospheric mass loading and crushing are neither ruled out nor favored conclusively by observations. In particular, $\dot{M}_\textrm{a,obs}$ is at least $\sim 10^4$ times $\dot{M}_\textrm{ff}$ for all four objects. However, without stronger observational constraints, we cannot conclude that mass loading or crushing is operating in these four objects, since quiescent accretion rates can be $\lesssim 10^3$ times lower than outburst rates \citep{hetal09b}. Note that it is enough for either one of the mass loading and crushing thresholds to be exceeded for the force-free spin-down formula to be modified. If either mass loading or crushing is activated, then the standard dipole spin-down formula should not be employed when calculating $B_*$ for these objects.} The magnetosphere is typically {mass-loaded} and/or crushed, so the braking torque is more like that described by \citet{gl79} for magnetised accreting stars (lever arm $R_\textrm{A}$) than the standard magnetic dipole torque (lever arm $R_\textrm{L}$). Indeed, the fact that the four accreting MSPs in Table 1 undergo recurring outbursts is circumstantial evidence that they are near magnetocentrifugal equilibrium, as is the low upper limit on the frequency derivative of {\sax} during outburst \citep{hp11}.

Near magnetocentrifugal equilibrium, a transiently accreting MSP is thought to exist in a `quasi-propeller regime', with $R_\textrm{A}$ fluctuating in the band $0.7R_\textrm{co}\lesssim R_\textrm{A}\lesssim 1.2 R_\textrm{co}$ \citep{rfs04,pbs06,hetal09,hgc11}. Magnetohydrodynamic simulations of the accretion disk in this state suggest that accretion onto the star either switches off completely or is inadequate to spin up the MSP, but the system does not enter the true propeller phase, where accreting matter is flung out by centrifugal forces \citep{rfs04,lrl05,pbs06}. In addition, \citet{rkl08} showed that there is a regime of unstable accretion, where the star accretes intermittently, and that the star can oscillate between the unstable and stable regimes. {If we assume that the star verges on the propeller phase, with $R_\textrm{A}\approx 1.2 R_\textrm{co}$ (Rappaport et al. 2004), analysis of the torques operating on the star has shown that the spin-down rate $\dot{\Omega}$ is approximated by}


\be I\dot{\Omega} = (1-\omega)\dot{M}_\textrm{a}(G\mstar R_\textrm{A})^{1/2},\label{torque}\ee
{where $\omega= (R_\textrm{A}/R_\textrm{co})^{3/2}$ is the fastness parameter, and $I$ is the moment of inertia of the accreting star \citep{gl79,rfs04,mp14}. We can then solve equation (\ref{torque}) simultaneously with the condition $R_\textrm{A}=1.2 R_\textrm{co}$ to find $B_*$ and $\dot{M}_\textrm{a}$}. This gives us not only an updated value of $B_*$, but also a consistency check on $\dot{M}_\textrm{a}$ for comparison with Table \ref{table2}.

We present $\dot{M}_\textrm{a}$ and $B_*$ for the four transient MSPs in Table \ref{table1}. We also list the spin frequency $\nu$, the quiescent spin frequency derivative $\dot{\nu}$, the `classically derived' surface magnetic field $B_{*, \textrm{classic}}=3.2\times 10^{19}(-\dot{\nu}/\nu^3)^{1/2}$ (assuming magnetic dipole braking with lever arm $R_\textrm{L}$), and the threshold accretion rates for the magnetospheric {mass loading} and crushing ($\dot{M}_\textrm{ff}$ and $\dot{M}_\textrm{L}$ respectively), assuming the surface magnetic field is indeed $B_{*,\textrm{classic}}$. The quiescent frequency derivatives are hard to pinpoint with certainty with existing X-ray timing data from the Rossi X-ray Timing Explorer (RXTE). We use the estimates given by \citet{hetal08}, \citet{p10}, \citet{pam10}, and \citet{retal11}.

\begingroup
\small
\begin{sidewaystable*}
\centering

  \caption{Modified magnetic moments of transient accreting MSPs with $\dot{\nu}$ measured during quiescence. The `classic' surface polar magnetic field strength $B_{*,\textrm{classic}}$ is obtained using the magnetic dipole spin-down formula $B_{\ast,\mathrm{classic}}=3.2\times 10^{19}(-\dot{\nu}/\nu^3)^{1/2}$ (lever arm $R_\textrm{L}$), whereas $B_*$ and the accretion rate $\dot{M}_\textrm{a}$ are calculated using the magnetised accretion torque (\ref{torque}). For comparison, we show the threshold accretion rates for {magnetospheric} mass loading (\S\ref{sec:qui2}) and crushing (\S\ref{sec:qui3}), assuming conservatively that $B_{*,\textrm{classic}}$ is true. All accretion rates are normalised to the Eddington rate $\dot{M}_\textrm{E}=3.1\times 10^{-9}$ $\msun$ yr$^{-1}$ ($\mstar=1.4$ $\msun$).}
  \begin{tabular}{lccccccc}
  \hline
    Name & $\nu$ & $\dot{\nu}$  & $B_{\ast,\mathrm{classic}}$  & $B_{*}$  & $\dot{M}_\textrm{a}/\dot{M}_{\textrm{E}}$ & $\dot{M}_\textrm{ff}/\dot{M}_{\textrm{E}}$ & $\dot{M}_\textrm{L}/\dot{M}_{\textrm{E}}$\\
         & (Hz) & ($10^{-15}$ Hz s$^{-1}$) & ($10^8$ G) & ($10^8$ G) & ($10^{-3}$) & ($10^{-9}$) & ($10^{-4}$)\\
\hline
IGR J00291+5934 & $ 598.89$ & $-3.0$ & $1.2$ & $0.7$ & $13$ & $90$ & $9.3$\\
XTE J1751$-$305 & 435.32 & $-5.5$ & $2.6$ & $1.4$ & $21$ & $100$ & $15$\\
SAX J1808.4$-$3658 & 401 & $-5.6\times 10^{-1}$ &  $0.9$ & $0.5$ & $2.1$ & $8.3$ & $1.4$\\
Swift J1756.9$-$2508 & 182 & $-2.0$ & $5.8$ & $2.0$ & $5.8$ & $9.8$ & $3.4$\\
\hline
\end{tabular}
\label{table1}
\end{sidewaystable*}
\endgroup

Table \ref{table1} leads to two main conclusions. First, the accretion rates inferred from (\ref{torque}) with $R_\textrm{A}=1.2R_\textrm{co}$ are higher than needed for poisoning and crushing to occur, so the use of equation (\ref{torque}) is justified a posteriori.\footnote{Using the newly derived $B_\ast$ yields lower {mass loading} and crushing thresholds.} Second, we find $B_*<B_{\ast,\mathrm{classic}}$ for all four objects, with $0.34\leqslant B_\ast/B_{\ast,\mathrm{classic}}\leqslant 0.58$. {From Tables 1 and 2, $\dot{M}_\textrm{a}$ for {\igr} and {\xte} are one order of magnitude higher than the quiescent $\dot{M}_\textrm{a,obs}$ estimate (Table 1). On the other hand, we find $\dot{M}_\textrm{a} \approx \dot{M}_\textrm{a,obs}$ for {\swi}. $\dot{M}_\textrm{a}$ for {\sax} is either approximately equal to $\dot{M}_\textrm{a,obs}$ or two orders of magnitude higher, depending on which estimate is used.}

How easy is it to adjust the system parameters to make $\dot{M}_\textrm{a}$ agree with $\dot{M}_\textrm{a,obs}$? For {\xte}, one would need to set $R_\textrm{A}=2.3R_\textrm{co}$ to give $\dot{M}_\textrm{a}/\dot{M}_\textrm{E}=1.9\times 10^{-3}$, to agree with $\dot{M}_\textrm{a,obs}$. For {\igr}, we need $R_\textrm{A}=2.7R_\textrm{co}$ to get $\dot{M}_\textrm{a}/\dot{M}_\textrm{E}=0.8\times 10^{-3}$. Nominally, such values of $R_\textrm{A}$ take the system far out of the quasi-propeller regime, and the star cannot be a transient accreting MSP, so we consider these values unlikely. If one uses these values of $R_\textrm{A}$ in (\ref{torque}), one finds $B_\ast=1.3\times 10^8$ G for {\xte} and $B_\ast=7.5\times 10^7$ G for {\igr}, which are close to $B_{\ast,\mathrm{classic}}$. This is pure coincidence, however, since the magnetic dipole and accretion torques are fundamentally different physically, e.g., they depend differently on $\mu$. It must be noted also that the presence of thermal components in the X-ray spectra of {\igr} and {\xte} during quiescence (taken by XMM-Newton) is uncertain \citep{jetal03,detal07,detal09,hetal09b}.

{As an alternative approach, we can estimate the quiescent $\dot{M}_\textrm{a}$ from the lowest X-ray flux with detected pulsations, like Mukherjee et al. (2015). Now, instead of solving for $\dot{M}_\textrm{a}$ and $B_*$ in (10) with $R_\textrm{A}=1.2 R_\textrm{co}$, we solve for $\omega$ (or, equivalently, $R_\textrm{A}$) and $B_*$ with the new values of $\dot{M}_\textrm{a}$ listed in the third column of Table 1. The results are shown in Table 3. We see again that $\dot{M}_\textrm{a,obs}$ is higher than the mass-loading threshold for all four objects and the crushing threshold for two out of four objects, but is borderline for IGR J00291$+$5934 and SAX 1808.4$-$3658. {Because the values derived by Mukherjee et al. (2015) correspond to upper limits of accretion rates, we cannot state unequivocally that this means the magnetospheres of the AMSPs are crushed or mass-loaded, we can only state that these scenarios (particularly mass loading) are not ruled out.} The accretion rate {upper limits} calculated by Mukherjee et al. (2015) gives values of $R_\textrm{A}/R_\textrm{co}$ that are fairly close to 1.2, i.e. the quasi-propeller regime (Rappaport et al. 2004), except for IGR J00291$+$5934, where we find $R_\textrm{A}/R_\textrm{co}=1.6$. We note also that our values of $B_*$ fall within the ranges inferred by Mukherjee et al. (2015).}

\begin{table*}
\centering
 \begin{minipage}{140mm}

  \caption{Modified magnetic moments of transient accreting MSPs with $\dot\nu$ measured during quiescence. The polar surface field strength $B_*$ and Alfv\'{e}n radius $R_\textrm{A}$ (given in terms of the corotation radius $R_\textrm{co}$) are calculated using (10) and the minimum pulsating fluxes given by Mukherjee et al. (2015). For comparison, we show the threshold accretion rates for magnetospheric mass loading (\S2, taking $\gamma = 10^6$) and crushing (\S3), assuming conservatively that $B_{*,\textrm{classic}}$ is true. All accretion rates are normalised to the Eddington rate $\dot{M}_\textrm{E}=3.1\times 10^{-9}$ $\msun$ yr$^{-1}$ ($\mstar=1.4$ $\msun$).}
  \begin{tabular}{lccccc}
  \hline
    Name & $\dot{M}_\textrm{a,obs}/\dot{M}_{\textrm{E}}$ & $R_\textrm{A}/R_\textrm{co}$ & $B_*$ & $\dot{M}_\textrm{ff}/\dot{M}_{\textrm{E}}$ & $\dot{M}_\textrm{L}/\dot{M}_{\textrm{E}}$\\
     & $(10^{-3})$ & & ($10^8$ G) & $(10^{-9})$ & $(10^{-4})$\\
\hline
IGR J00291+5934 & 3.2 & 1.6 & 0.63 & 90 & 9.3\\
XTE J1751$-$305 & 63 & 1.1 & 1.93 & 100 & 15\\
SAX J1808.4$-$3658 & 1.1 & 1.3 & 0.42 & 8.3 & $1.4$\\
Swift J1756.9$-$2508 & 41 & 1.0 & 4.02 & 9.8 & $3.4$\\
\hline
\end{tabular}
\end{minipage}
\label{muk}
\end{table*}









\section{Discussion
 \label{sec:qui7}}

In this paper, we argue that the standard magnetic dipole spin-down formula overestimates the field strengths of transient accreting MSPs with $\dot{\nu}$ measured during quiescence, in particular {\igr}, {\xte}, {\sax}, and {\swi}. {Current observational estimates for the residual accretion rates can only set (relatively lax) upper limits, which neither rule out nor favor conclusively magnetospheric mass loading or crushing. If it transpires that residual accretion during quiescence does mass load or crush the magnetosphere, then the magnetosphere is distorted away from a force-free configuration and the torque lever arm is shortened from $R_\textrm{L}$ to $R_\textrm{A}$.} Under these conditions, spin down during quiescence occurs due to a Ghosh-Lamb-like magnetized accretion torque with $R_\textrm{A}>R_\textrm{co}$ in the quasi-propeller regime \citep{rfs04,pbs06,hetal09,hgc11}. Assuming $R_\textrm{A}=1.2R_\textrm{co}$, by way of illustration, we find $B_*<B_{\ast,\mathrm{classic}}$ for all four MSPs (Table \ref{table1}), i.e., the standard spin-down formula overestimates $B_\ast$ by a factor of $\lesssim 3$. The accretion rate inferred thus is consistent with estimates {(averaged over outbursts and quiescence)} for {\sax} and {\swi}, and higher by about an order of magnitude for {\igr} and {\xte}. {Compared to $\dot{M}_\textrm{a,obs}$ estimates of Mukherjee et al. (2015), our values are slightly larger for {\sax} and {\swi}, one order of magnitude higher for {\igr}, and three times smaller for {\xte}.} The inferred $B_\ast$ values remain consistent with recycling-related scenarios of $\mu$ reduction, such as polar magnetic burial \citep{pm04,petal11b}. {Note that we do not derive a new spin-down formula rigorously; our goal is simply to bring attention to the effects of residual accretion on the spin down of a transient accreting MSP. Nevertheless, equation (10) is a good approximation to $\dot{\Omega}$ for $R_\textrm{lever}\approx R_\textrm{A} < R_\textrm{L}$, when inertial effects are important.} We stress again that systematic uncertainties affect both the theoretical and observational facets of the problem.

One may ask why there is accretion at all during a transient accreting MSP's quiescence. For $R_\textrm{A}\approx R_\textrm{co}$ (as discussed in \S\ref{sec:qui6}, for example), propellered matter is not likely to gain sufficient speed to escape the system \citep{st93,rfs04,das10}. \citet{rfs04} showed that the angular momentum given to the disk by the neutron star, via the magnetosphere, is transported outwards, so that matter at the inner radius of the disk does not acquire enough speed to escape, leading to a buildup of matter near $R_\textrm{co}$. Effectively, the inner disk radius is located just inside $R_\textrm{co}$, even for $R_\textrm{A}>R_\textrm{co}$, even for small accretion rates $\dot{M}_\textrm{a}\lesssim 10^{-11}$ $\msun$ yr$^{-1}$ \citep{rfs04,kl07}. Furthermore, \citet{lrb-k99} suggested that the effective Alfv\'{e}n radius, as opposed to the nominal Alfv\'{e}n radius given by (\ref{eqra}), depends on $\Omega$ as well, and wanders around $R_\textrm{co}$ stochastically or chaotically, triggered by small variations in $\dot{M}_\textrm{a}$ or magnetic field configuration. In fact, even in the propeller regime, some quasiperiodic accretion still occurs \citep{retal04}. Thus, the disk-magnetosphere interaction is never enough to halt accretion completely, and there is always some matter accreting inside $R_\textrm{A}$. In addition, as mentioned in \S\ref{sec:qui5}, larger, neutral particles in the disk can undergo collisional or Yarkovsky migration into the magnetosphere \citep{cor08}. In a slightly different physical regime than the one we discuss here, \citet{rl06} showed that the misalignment angle between the magnetic axis and the rotational axis affects how much matter can migrate into a solar-type protostar's magnetosphere. Repeating their simulation for an accreting MSP may yield interesting results.

Incidentally, other mechanisms may also disrupt the magnetosphere and modify the braking torque. For example, the pulsar is encased in a conducting cage of accreting plasma. Even if the magnetosphere is not crushed (i.e., $R_\textrm{A}>R_\textrm{L}$), the cage reflects the low-frequency, large-amplitude electromagnetic or magnetohydrodynamic wave in the Poynting-flux-dominated wind inside the accretion shock \citep{kk80,c90,mm96,sks05,ak13}.\footnote{If the wind transitions from a Poynting- to a kinetic-dominated outflow inside the termination shock, the shock has the capacity to emit strongly in X-rays \citep{kc84,mm96,cha07}. This possibility and its implications for pulsar wind physics deserve further study in the context of accreting millisecond pulsars like SAX J1808.4$-$3658.} This is analogous to sealing an antenna inside a partially reflecting, conducting box. If the cage has large inertia, the reflected wave bounces back onto the pulsar, modifying the spin-down torque away from its standard magnetic dipole form.

In addition to residual accretion from a remnant disk, Bondi-Hoyle accretion also occurs, as the pulsar travels through the interstellar medium. This serves as an important sanity check on the quenching mechanisms described in \S\ref{sec:qui2} and \S\ref{sec:qui3}, because we know that isolated MSPs are routinely detected as radio sources and often show other evidence for a functioning pulsar machine, e.g.\ the relativistic wind in the H$\alpha$ bow shock nebula around PSR J0437$-$4715 \citep{bel95}. Hence, vacuum gap poisoning {(although not necessarily magnetospheric crushing or mass loading)} by accretion of the interstellar medium {is} ruled out observationally in such objects. The Bondi-Hoyle accretion rate for a pulsar with speed $V_\ast \ll V_{\rm th}$, where $V_{\rm th}$ is the thermal speed in the interstellar medium, and $V_\ast$ is the sum of orbital (binary) and translational (kick) velocity components, is given by

\begin{eqnarray}
 \frac{\dot{M}_{\rm ISM}}{\dot{M}_{\rm E}}
 & = &
 n_{\rm ISM} \rstar  \sigma_{\rm T}
 \left(
  \frac{V_\ast}{c}
 \right)^{-3}
 \left(
  \frac{GM_\ast}{\rstar c^2}
 \right)
\label{eq:qui21}
 \\
 & = &
 3.8\times 10^{-9}
 \left(
  \frac{n_{\rm ISM}}{1\, {\rm cm^{-3}}}
 \right)
 \left(
  \frac{V_\ast}{10^2\,{\rm km\,s^{-1}}}
 \right)^{-3}~,
\label{eq:qui22}
\end{eqnarray}
where $n_{\rm ISM}$ is the proton number density in the interstellar medium. Comparing (\ref{eq:qui22}) with (\ref{eq:qui4}) and (\ref{eq:qui7}), we see that Bondi-Hoyle accretion is unlikely to crush the magnetosphere. Interestingly, it is borderline for poisoning certain objects; see also \citet{cor08} and references therein. {We stress again that gap poisoning affects the radio emission, but it does not affect the spin-down torque (except indirectly through $\gamma$ is equations (5) and (6).}

{Recently, Mukherjee et al. (2015) estimated the minimum and maximum surface field strengths of 14 accreting MSPs as follows. The minimum $B_\ast$ is found by setting the disk truncation radius $(\approx R_\textrm{A}$ up to a boundary layer correction factor of order unity) equal to $R_\ast$, with $\dot{M}_\textrm{a}$ given by the maximum pulsating X-ray flux in equation (7). The maximum $B_\ast$ is found by setting the truncation radius equal to $R_\textrm{co}$, with $\dot{M}_\textrm{a}$ inferred from the minimum pulsating X-ray flux. This approach resembles ours leading to equation (10), except that we match $R_\textrm{A}=R_\textrm{A}(\dot{M}_\textrm{a},B_\ast)$ to some radius just outside $R_\textrm{co}$ and assume that the MSP hovers between the accreting and propeller regimes (Hartman, Galloway, \& Chakrabarty 2011; D'Angelo \& Spruit 2012). Our calculated values of $B_*$ fall within the ranges obtained by Mukherjee et al. (2015) and are lower than those obtained purely from quiescent spin down, as Mukherjee et al. (2015) found independently. This agreement reinforces our argument that quiescent spin down cannot be used na\"{i}vely to estimate $B_*$. Note that only one mechanism (magnetospheric mass loading or crushing) needs to be activated to modify the spin-down torque away from classical dipole expectation.}

In the future, it would be worth looking for direct observational signatures of {non-force-free magnetospheres} in transient accreting MSPs. However, it remains to be seen whether such signatures can be interpreted unambiguously. For example, there have already been detections, during quiescence, of sinusoidal modulations of the optical flux from the companion of {\sax} \citep{del08,wetal09}, whose photometric maxima occur whenever the neutron star is between
the companion and the observer. \citet{hom01} originally interpreted the modulations as emission from a non-irradiated accretion disk truncated at the corotation radius. More recently, however, it has been argued that the neutron star switches on during quiescence as a rotation-powered pulsar, whose relativistic wind irradiates one hemisphere of the companion \citep{bur03,dis03}, {although such irradiation (by a Poynting-flux-dominated outflow) occurs whether or not the star is a pulsar}. By analysing the spin distributions of MSPs, \citet{petal14} found that there is a $90\%$ probability that accreting MSPs and eclipsing rotation-powered MSPs [rotation-powered MSPs that show irregular eclipses in their radio emission, caused by matter irradiated away from the companion by the pulsar \citep{r13}] belong to the same population. {A number of MSPs in close binary systems are active as radio pulsars which
emit winds that prevent the formation of accretion disks (Roberts 2013). Recently, one of these systems (IGR J18245$-$2452) has been observed to behave as an accreting MSP during an X-ray outburst, and as a radio pulsar during quiescence (Papitto et al.
2013), indicating the tight link between AMSPs and radio pulsars.} However, searches for radio pulsations in the four MSPs discussed in this paper have been carried out without success \citep{iac10,petal14}.\footnote{No radio emission, pulsed or otherwise, has been detected during outbursts either \citep{tud08}.} There are many reasons why this might be so, e.g., beaming. A neutron star can act like a rotation-powered pulsar electrodynamically (with a Poynting-flux-dominated wind flowing out from the light cylinder and carrying most of the spin-down luminosity) without being a magnetospheric radiation source, cf. \citet{luo07}. By the same token, a neutron star can heat its companion without switching on as a rotation-powered pulsar; an accretion-dominated magnetosphere carries an outward-directed Poynting flux, even when the Goldreich-Julian current system is disrupted. Hence the optical modulations observed from {\sax} have several valid interpretations. More multiwavelength studies are needed to clarify the situation.

\acknowledgments
This research was supported by an Australian Research Council Discovery Project grant (DP110103347). We thank the two anonymous referees for their comments, which substantially improved the paper. AMe thanks Deepto Chakrabarty and Scott Hughes for stimulating discussions during a sabbatical visit to the Massachusetts Institute of Technology, where the ideas behind the paper were first formulated.

\bibliographystyle{mn2e}
\bibliography{amsp}

\end{document}